# Purified graphite surface and vacancy states: undercoordination-induced quantum trap depression and lone π-electron polarization

Chang Q Sun,[1,4] Yanguang Nie,[1] Jisheng Pan,[2] Weitao Zheng[3]

[1]*School of Electrical and Electronic Engineering, Nanyang Technological University, Singapore 639798;* [2]*Institute of Materials Research and Engineering, A*Star, Singapore 117602;* [3]*School of Materials Science, Jilin University, Changchun 130012, China;*[4]*Faculty of Materials, Photoelectronics and Physics, Xiangtan University, Changsha 400073, China*

Ecqsun@ntu.edu.sg

We present a simple approach for purifying graphite surface and vacancy states using an angle-resolved x-ray photoelectron residual spectroscopy (XPS). Complementing the discoveries of Ugeda et al [*Phys Rev Lett 104*, 096804 (2010)], outcomes conform the BOLS theory [Sun, *Prog Solid State Chem 35*, 1-159 (2007)] expectation and the recent findings that the shorter and stronger bonds between undercoordinated atoms induce local strain and quantum potential depression with an association of local densification of energy and core electrons. The shorter atomic distance and the densely and deeply trapped bonding and core charges polarize in turn the unpaired π-electrons nearby vacancy, giving rise to the high protrusions and the Dirac $E_F$ states as observed. The quantum trap depression and the screening due to the polarized $E_F$ states split the crystal potential and hence the extra XPRS C 1s states.

Undercoordinated atoms at sites surrounding atomic vacancies, chain ends, defects, cavities, terrace edges, and the skins of nanostructures demonstrate properties that cannot be seen even from a flat surface. Scanning tunneling microscopy/spectroscopy (STM/S) measurements revealed extraordinarily high protrusions associated with a resonant peak of Dirac states centered around $E_F$ for atoms surrounding an atomic vacancy at graphite surface opposing to atoms at the clean surface.[1-3] Presented also at the graphene nanoribbon (GNR) edge, the Dirac states move further to higher energy when the STM tip moves closer to the edge. [4-7] The GNR interior does not show such significance. [6] These observations indicate clearly charge polarization taking place nearby vacancy. These edge states are almost massless with a group velocity of 1/300 that of light travelling in vacuum and with the abnormal magnetism and fractional quantum Hall effect,[3, 8-9] while the mobility of the vacancy states are limited.[1] As the STM/S collects information in the vicinity of $E_F$ at the atomic scale from a surface, more information regarding the origin for the protrusions and the Dirac states and mobility limitation at the vacancy and the correlation between the $E_F$ states and the core charges in deeper energy bands is highly desired.

In the X-ray photoelectron spectroscopy (XPS) measurement of graphite, it has been observed that the C1s peak is broadened upon vacancy generation induced by $Ar^+$ ion spraying.[10] The spectrum is also broadened by increasing the angle between the photoelectron beam and the surface normal, or called emission angle.[11-12] Large-angle XPS collects more information from undercoordinated surface atoms.[10]

The STM/S and XPS findings are indeed fascinating and inspiring. In order to correlate the STM/S and XPS observations, we conducted the angle-resolved x-ray photoelectron residual spectroscopy (XPRS) measurements of graphite surface with and without artificial vacancies at room temperature. After proper background subtraction and area normalization under the guideline of spectral area conservation, the spectra collected at 75° emission angle were subtracted by the one collected from the clean surface at the smallest (25°) emission angle. The rule of area conservation means that the integration of each spectrum is proportional one to another because of the effect of scattering. Such an XPRS process purifies the surface and the vacancy states as the XPRS filters out the bulk information. Speranza and Minati[13] found that using the same Al $K_{\alpha}$ source the X-ray penetration depth in graphite decreases from 8.7 to 0.7 nm when the emission angle is increased from 0 to 85°.

The principles are very simple. Firstly, the core-level energy shift from that of an isolated atom is dominated by the crystal potential in the Hamiltonian. The eigen wave functions for the core electrons remain unperturbed by coordination reduction as these electrons are strongly localized.[14] Any perturbation in the Hamiltonian of an extended bulk solid will lead to the core level to shift further from that of the bulk. The direction of the shift depends on the perturbation to the potential. The perturbation includes bond contraction, bond nature alteration, charge polarization, and other external stimuli. As the crystal potential at equilibrium corresponds to the bond length and bond energy, the core level shift is proportional to the bond energy. Secondly, according to Pauling[15] and Goldschmidt,[16] bonds become shorter and stronger when the coordination number (z) of an atom is reduced. This spontaneous process of bond contraction will lower the bond potential energy and hence the core level energy, leading to the quantum trap depression, with an association of local densification of charges and energies.[9] Thirdly, the shorter atomic distance and the deeply and densely trapped core charges will polarize the weakly bound nonbonding charge such as the cases of the half-filled s-orbit of Rh surface adatoms[17] and the unpaired π-bond electrons near the vacancy in graphite or graphene, as widely observed.[1, 5-7, 9] The inhomogeneous and localized polarization (P) of the unpaired charge will partially in turn screen and split the crystal potential and hence generate extra states in the core bands.

STM/S and XPRS are correlated and complement each other. The former collects information about the polarized states near $E_F$ and images the surface dipoles as protrusions at the atomic scale;[18] the latter collects statistic information about the deeper core bands with the trapped T and the screened P states from the surface skin limited to three atomic layers as finger prints of that happened at the surface. Their combination could provide comprehensive information about the energetic behavior of the bonding and the nonbonding electrons and their interdependence at sites surrounding undercoordinated atoms.

Analytically, the core level shift can be formulated by the combination of band theory [14] and the bond order-length-strength (BOLS) correlation mechanism.[19] The single-body Hamiltonian is perturbed by the shorter and stronger bonds and the screening effect of polarization, denoted with $\Delta_H$ :

$$H(\Delta_H) = -\frac{\hbar\nabla^2}{2m} + V_{atom}(r) + V_{cry}(r)[1+\Delta_H],$$

where

$$1+\Delta_H = \begin{cases} C_z^{-m} = E_z / E_0 & (Trap\ depression) \\ p = (E_{1s}(p) - E_{1s}(0))/(E_{1s}(12) - E_{1s}(0)) & (Polarization) \end{cases}$$
$$C_z = d_z / d_0 = 2/\{1+\exp[(12-z)/(8z)]\}$$

(1)

$E_{1s}$ (x) represents the peak energy of the z or P component in the XPRS. $C_z$ is the Goldschmidt-Pauling coefficient of bond contraction. The *p* is the coefficient of polarization to be determined from the XPRS. $\Delta E_{1s}(12)$ is the bulk shift of diamond with an effective z vale of 12 rather than 4 because the diamond is an interlock of two fcc unit cells.

The BOLS theory reproduction [9, 20] of the elastic modulus[21-22] and the melting point[23] of carbon nanotubes, and the C1s core level shift of carbon allotropes[24-25] has revealed consistently that the C-C bonds between two-coordinated atoms contract by 30% from 0.154 nm to 0.107 nm and the bond energy increase by 150% with respect to those of diamond, giving a generalized form for the z-resolved C1s binding energy shift [25] with the bond nature indicator m = 2.56,

$$E_{1s}(z) = E_{1s}(0) + [E_{1s}(12) - E_{1s}(0)]\begin{cases} C_z^{-2.56} \\ p \end{cases} = 282.57 + 1.32\begin{cases} C_z^{-2.56} \\ p \end{cases} (eV) \tag{2}$$

The experimental discovery[7] that the minimal energy (7.5 eV/bond) required for breaking a 2-coordinated carbon atom near vacancy is 32% higher than that (5.67eV/bond) required for that of a 3-coordinated carbon atom in graphene provides more evidence for the BOLS prediction that the broken bonds do enhance the neighboring bond strength. The findings of gold cluster surface bond contraction [26] and Nb[27] and Ta[28] surface relaxation also conform the BOLS expectation.

The XPS experiments were conducted using the Sigma Probe Instrument (Thermal Scientific) with monochromatic Al $K_\alpha$(1486.6 eV) as the x-ray source. The XPS was calibrated using pure gold, silver, and copper standard samples by setting the Au-4$f_{7/2}$, Ag-3$d_{5/2}$ at binding energy of 83.98 ± 0.02 eV, 368.26 ± 0.02 eV, respectively. The emission angle was varied within the range of 25° and 75°. The highly oriented pyrolytic graphite (HOPG) was cleaved using the adhesive tape in air, and then, transferred rapidly into the XPS chamber. In order to examine the effect of vacancy generation, the surface was sputtered using $Ar^+$ ions with 0.5, 1.0, 3.0 KeV energy incident along the HOPG surface normal. The spray dose was controlled by the sample current and the duration of sputtering.

The collected raw data from the vacancy surfaces at 50° and from the vacancy-free surfaces at different emission angles, as shown in Figure 1a and b, show the broadened and attenuated spectral features. These spectra were normalized after background subtraction using the Shelly method. The spectrum collected at the smallest (25°) emission angle from the vacancy-free surface was used as a reference for the XPRS subtraction, as the x-ray beam at the smallest angle collects bulk-dominated information. Such an XPRS will purify the information dominated by electrons in the outermost three atomic layers or estimated 1.0 nm dpeth.[13] The additional advantage of such a XPRS process is the minimization of the influence by extrinsic factors such as the background uncertainty and the "initial-final states" effect that exists throughout the course of measurements.

Figure 1c and d show the evolution of the XPRS with the variation of the vacancy density and the emission angle, respectively. The area above the E-axis represents states gain and the area below the E-axis the states loss under the given conditions. Core charges with energy at the valley will go to the trapped or to the polarized states upon vacancy formation. The net gain should be zero because of the rule of spectral area conservation. For the vacancy-free surface, only one trapped peak presents and the peak position shifts gradually from energy corresponding to z ~ 4 to energy of z ~ 3.2, as the emission angle increases from 35° to 75°. The z value of 5.335 for graphite was obtained by applying the Goldschmidt-Pauling coefficient to the known C-C bond length of graphite (0.142 nm) and that of diamond (0.154 nm) with the effective z of 12. The presence of the T peak in both cases verifies the expectation of quantum trap depression; the presence of P states to vacancy due to the polarization of the unpaired π electrons of the dangling bond. It is seen from (d) that the extent and energy shift of the polarization is proportional to the extent of T states. The presence of both the T and the P states to the vacancy XPRS profiles confirms the crystal potential screening and splitting caused by less atomic coordination.

Figure 3 summarizes our findings of the purified XPRS collected at 75° emission angle from surfaces with and without vacancies. The effect atomic z of the graphite skin is ~3.2 and the z for vacancy neighbors is ~2.5. These two values may vary with defect density. The vacancy P states are centered at 283.63 eV, 0.31 eV above that (283.94 eV) of bulk diamond. Therefore, the *p* = (283.63-282.57)/1.32 = 1.06/1.32= 0.80, which means that screened potential is 20% shallower than that in diamond, while the vacancy trapping potential is $C_{z=2.5}^{-2.56}-1=0.97$ times deeper. With the obtained z values, one can estimate the local bond length ($C_z$), bond energy ($C_z^{-2.56}$), atomic cohesive energy $\left(zC_z^{-2.56}/12\right)$, and charge density $\left(C_z^{-3}\right)$ localized in the respective region.

Findings clarify thus the origin for the vacancy charge polarization and mobility limitation and their correlation with the C1s band P and T states. It is expected that the unusual catalytic reactivity and magnetism associated with the undercoordination of carbon arises from the polarized electrons that add impurity states above the valence band.[29]

In summary, the XPRS findings do complement the STM/S observations from the HOPG vacancy surface, which conform the BOLS theory prediction and the associated findings that the shorter and stronger bonds between undercoordinated atoms induce local strain and quantum trap depression with an association of local densification of charge and energy. This effect provides perturbation to the Hamiltonian that dominates globally the shifts of all levels to deeper energies. The densely and deeply trapped bonding charge will in turn polarize the weakly bound nonbonding (unpaired π-bond) electrons, leading to the STM/S mapped protrusions and the Dirac $E_F$ states. The polarization of the nonbonding electrons will screen partially in turn the crystal potential, giving rise to the P states in the XPRS profile. The effect of trap is significant though polarization is unapparent for atoms at the vacancy-free surface because of the lacking of the lone π-bond electrons, as detected using STM/S.

Findings herewith clarify the energy correlation between different bands and the driving force for the polarization and mobility limitation due to the artificial undercoordination of carbon atom at the surface and around the vacancy. XPRS is able to identify statistically the T states and the effect of polarization due to crystal potential screening and splitting; STM/S observes directly the atomic protrusions and the polarized $E_F$ states. A combination of the two methods with the theories of Goldschmidt-Pauling bond contraction and BOLS correlation is more revealing than using STM/S alone in examining the energetic and electronic behavior of the undercoordinated systems and their unusual properties that are determined by the interatomic bonding and charge distribution.

Figure Captions

**Figure 1** Schematic illustration of the atomic undercoordination induced Goldschmidt-Pauling bond contraction ($d_z < d_0$) and the associated energy and bonding charge quantum trapping (T), which polarize (P) the unpaired π-electrons generating the STM protrusions and the $E_F$ states. The polarized charge will partially in turn screen and split the crystal potential that determines the core level shift intrinsically.

Figure 2 The raw XPS spectra collected from (a) vacancy-free HOPG surface at different emission angles and (b) vacancy surface at 50° at different doses of $Ar^+$ spray. The (c) angle-resolved vacancy-free XPRS show only the trapped states (T) corresponding to atomic CN of ~3.2. The (c) vacancy-density-resolved XPRS spectra show both the trapped (T) at CN of ~2.5 and polarized states (P) centered at 283.63 eV.

Figure 3 Comparison of the purified XPRS C 1s spectra collected at 75° from the surface with ($9\times10^{14}$ $cm^{-2}$ dosed $Ar^+$ spray) and without vacancies. The XPRS P states correspond to the STM/S protrusions and the Dirac $E_F$ states and they are originated from the polarization of the weakly bound lone electrons by the densely and deeply trapped core and bond charges (T states). At a vacancy-free surface, neither STM/S protrusions nor the P states present though the trap states remain.

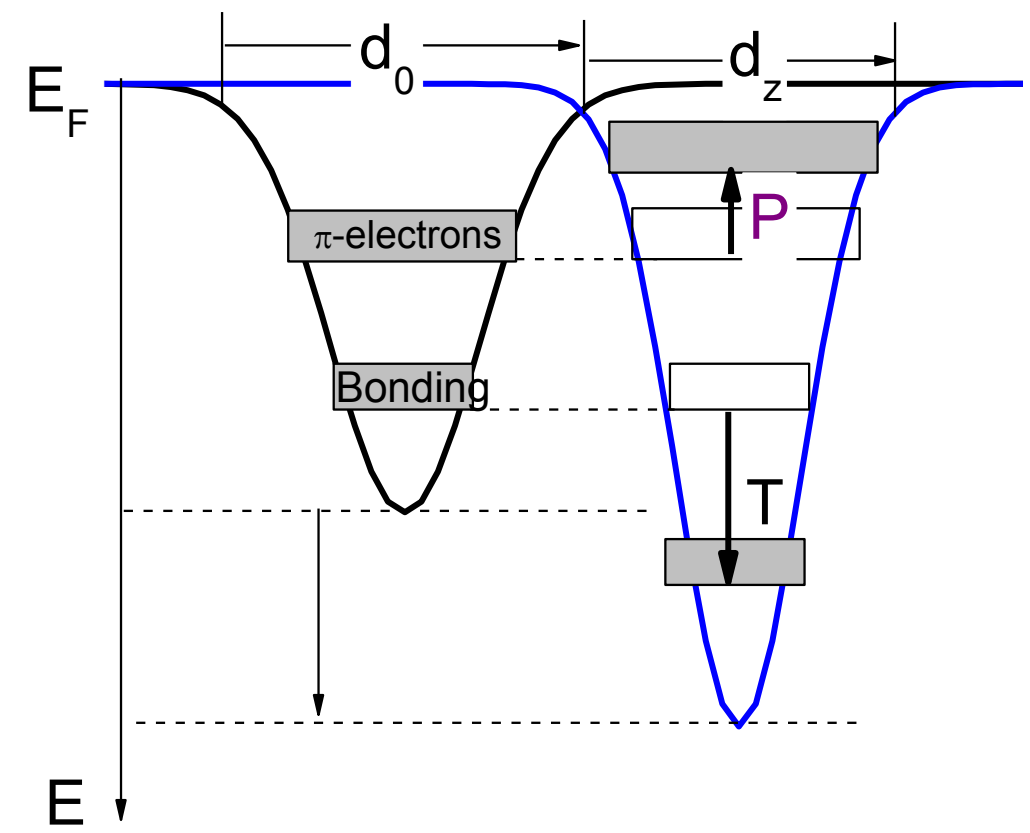

Figure 1

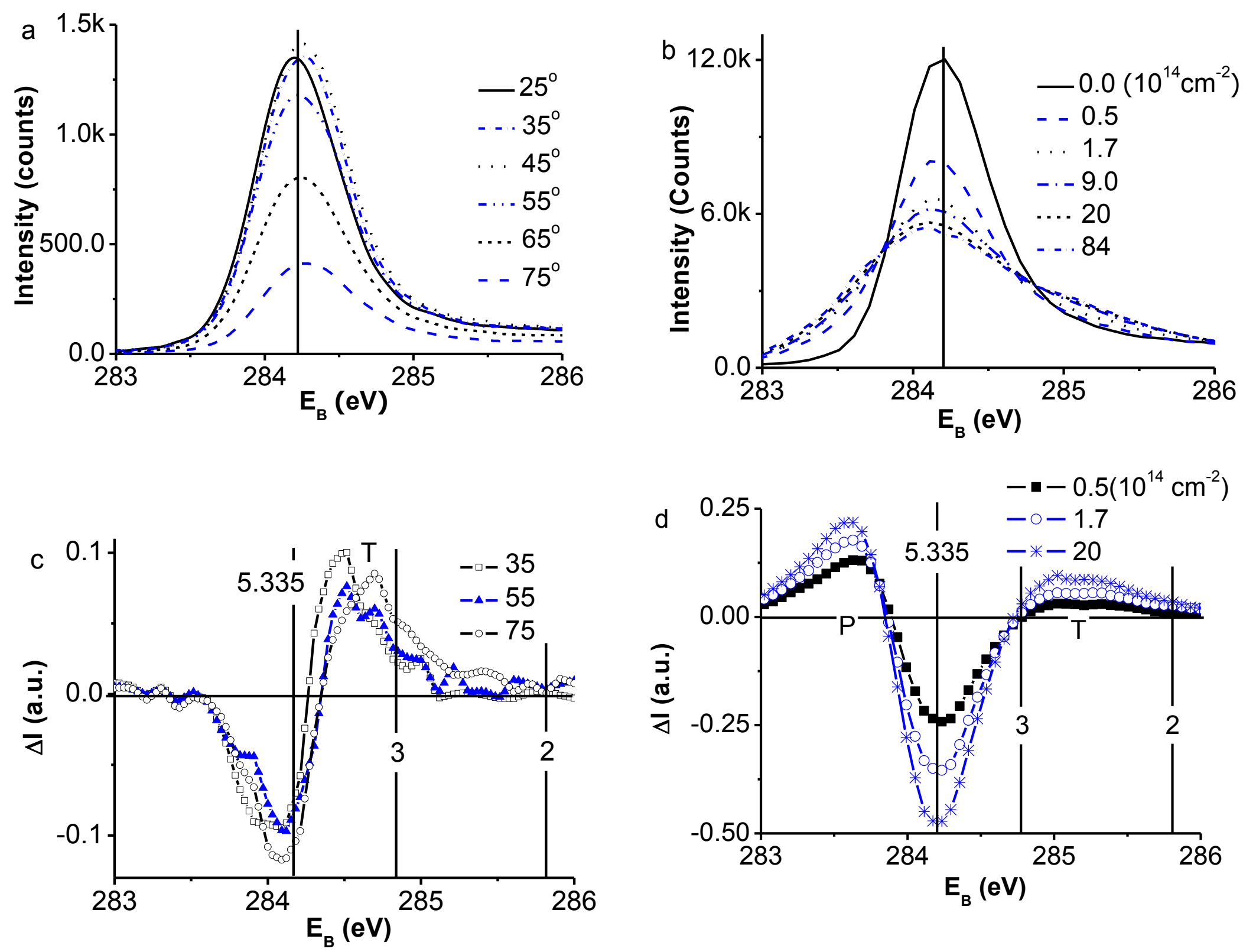

Figure 2

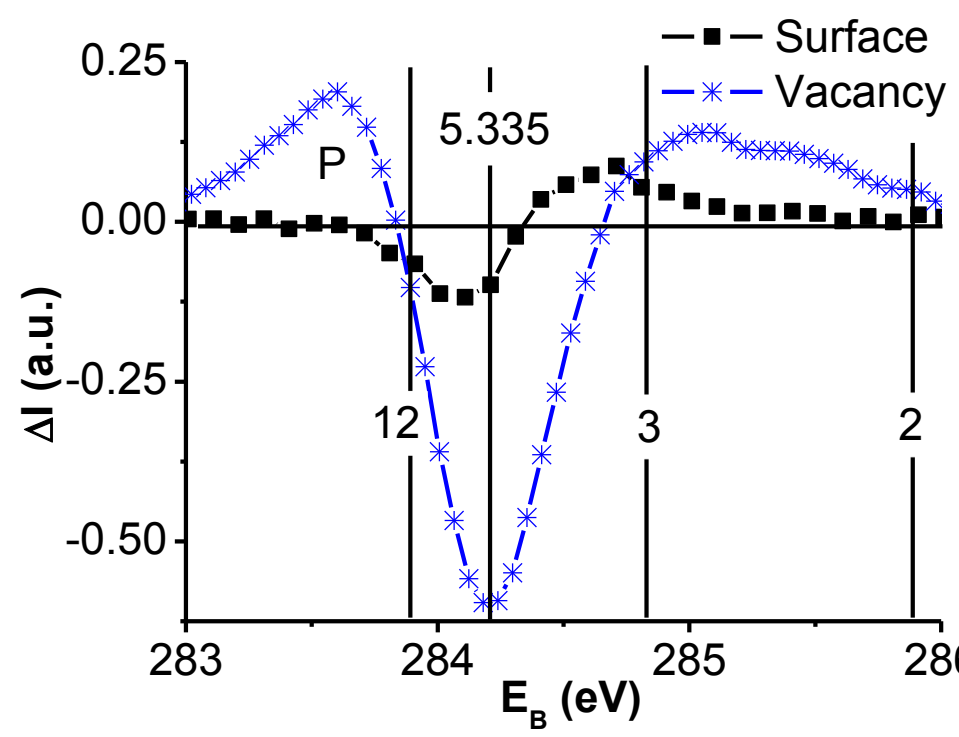

Figure 3

1. M. M. Ugeda, I. Brihuega, F. Guinea and J. M. Gómez-Rodríguez, Phys. Rev. Lett. **104**, 096804 (2010).
2. T. Matsui, H. Kambara, Y. Niimi, K. Tagami, M. Tsukada and H. Fukuyama, Phys. Rev. Lett. **94**, 226403 (2005).
3. G. Li and E. Y. Andrei, Nature Physics **3**, 623 (2007).
4. Y. Niimi, H. Kambara and H. Fukuyama, Phys. Rev. Lett. **102**, 026803 (2009).
5. Y. Niimi, T. Matsui, H. Kambara, K. Tagami, M. Tsukada and H. Fukuyama, Phys Rev B **73**, 085421 (2006).
6. T. Enoki, Y. Kobayashi and K. I. Fukui, Int. Rev. Phys. Chem. **26**, 609 (2007).
7. C. O. Girit, J. C. Meyer, R. Erni, M. D. Rossell, C. Kisielowski, L. Yang, C. H. Park, M. F. Crommie, M. L. Cohen, S. G. Louie and A. Zettl, Science **323**, 1705 (2009).
8. K. I. Bolotin, F. Ghahari, M. D. Shulman, H. L. Stormer and P. Kim, Nature **462**, 196 (2009).
9. C. Q. Sun, S. Y. Fu and Y. G. Nie, J Chem Phys C **112**, 18927 (2008).
10. T. Balasubramanian, J. N. Andersen and L. Wallden, Phys Rev B **64, 205420** (2001).
11. K. Y. Wu, W. Y. Chen, J. Hwang, H. K. Wei, C. S. Kou, C. Y. Lee, Y. L. Liu and H. Y. Huang, Applied Physics a-Materials Science & Processing **95**, 707 (2009).
12. D. Q. Yang and E. Sacher, Surf. Sci. **504**, 125 (2002).
13. G. Speranza and L. Minati, Surf. Sci. **600**, 4438 (2006).
14. M. A. Omar, *Elementary Solid State Physics: Principles and Applications*. (Addison-Wesley, New York, 1993).
15. L. Pauling, J. Am. Chem. Soc. **69**, 542 (1947).
16. V. M. Goldschmidt, Berichte Der Deutschen Chemischen Gesellschaft **60**, 1263 (1927).
17. C. Q. Sun, Y. Wang, Y. G. Nie, Y. Sun, J. S. Pan, L. K. Pan and Z. Sun, J Chem Phys C **113**, 21889 (2009).
18. C. Q. Sun, Prog. Mater Sci. **48**, 521 (2003).
19. C. Q. Sun, Phys Rev B **69, 045105** (2004).

20. C. Q. Sun, H. L. Bai, B. K. Tay, S. Li and E. Y. Jiang, J. Phys. Chem. B **107**, 7544 (2003).
21. M. R. Falvo, G. J. Clary, R. M. Taylor, V. Chi, F. P. Brooks, S. Washburn and R. Superfine, Nature **389**, 582 (1997).
22. E. W. Wong, P. E. Sheehan and C. M. Lieber, Science **277**, 1971 (1997).
23. B. An, S. Fukuyama, K. Yokogawa and M. Yoshimura, Jpn. J. Appl. Phys. **37**, 3809 (1998).
24. K. Ki-jeong, L. Hangil, C. Jae-Hyun, Y. Young-Sang, C. Junghun, L. Hankoo, K. Tai-Hee, M. C. Jung, H. J. Shin, L. Hu-Jong, K. Sehun and K. Bongsoo, Adv. Mater. **20**, 3589 (2008).
25. C. Q. Sun, Y. Sun, Y. G. Nie, Y. Wang, J. S. Pan, G. Ouyang, L. K. Pan and Z. Sun, J Chem Phys C **113**, 16464 (2009).
26. W. J. Huang, R. Sun, J. Tao, L. D. Menard, R. G. Nuzzo and J. M. Zuo, Nat. Mater. **7**, 308 (2008).
27. B. S. Fang, W. S. Lo, T. S. Chien, T. C. Leung, C. Y. Lue, C. T. Chan and K. M. Ho, Phys Rev B **50**, 11093 (1994).
28. R. A. Bartynski, D. Heskett, K. Garrison, G. Watson, D. M. Zehner, W. N. Mei, S. Y. Tong and X. Pan, J. Vac. Sci. Technol., A **7**, 1931 (1989).
29. J. M. D. Coey, Curr. Opin. Solid State Mater. Sci. **10**, 83 (2006).